\documentclass{elsart}
\usepackage{natbib}
\usepackage{epsfig}
\usepackage{amssymb}
\newcommand{\be}{\begin{equation}}      
\newcommand{\ee}{\end{equation}}      
      
\newcommand{\bef}{\begin{figure}}      
\newcommand{\eef}{\end{figure}}     
\def\spose#1{\hbox to 0pt{#1\hss}}      
\def\ltapprox{\mathrel{\spose{\lower 3pt\hbox{$\mathchar"218$}}      
 \raise 2.0pt\hbox{$\mathchar"13C$}}}      
\def\gtapprox{\mathrel{\spose{\lower 3pt\hbox{$\mathchar"218$}}      
 \raise 2.0pt\hbox{$\mathchar"13E$}}}      
\def\inapprox{\mathrel{\spose{\lower 3pt\hbox{$\mathchar"218$}}      
 \raise 2.0pt\hbox{$\mathchar"232$}}}
\begin{document}

\begin{frontmatter}

  \title{Complexity in cosmic   structures} 
  \author{Francesco Sylos Labini} \ead{sylos@th.u-psud.fr}
  \address{Laboratoire de Physique Th\'eorique Batiment 210
  Universit\'e de Paris XI, 91405 Orsay Cedex, France}
  \author{Andrea Gabrielli} \ead{Andrea.Gabrielli@roma1.infn.it}
  \address{``Enrico Fermi'' Center, 
Via Panisperna 89A, Compendio del Viminale, 
00184, Rome, Italy}

\begin{abstract}
We discuss correlation properties of a general mass density field
introducing a classification of structures based on their complexity.
Standard cosmological models for primordial mass fluctuations are
characterized by a sort of large-scale {\em stochastic} order, that we
call {\em super-homogeneity} to highlight the fact that mass
fluctuations increase as a function of scale in the slowest
possible way for any stochastic mass field.  On the other hand the
galaxy spatial distribution show complex structures with a high degree
of inhomogeneity and fractal-like spatial correlations up to some
relevant cosmological scale. The theoretical problem of cosmological
structure formation should then explain the growth of strongly
correlated and non-linear structures from the very uniform field of
density fluctuations given as standard initial condition.
\end{abstract}

\end{frontmatter}


Cosmic structures represent a very interesting playground for the
methods of statistical physics and self-organization of complex
structures. There are two main areas. The first deals with the very
irregular spatial structures developed by the clustering of galaxies.
The second broad area is provided by the cosmic microwave background
radiation (CMBR) which is extremely smooth apart from very
small-amplitude fluctuations. A general theory should link these two
observations which appear quite different.  Many new data for both
areas are now available and much more are expected in the near future,
creating big expectations, interest and animating challenges. Indeed,
on the observational side, the data in cosmology have been growing
exponentially in the last ten years, and in the coming decade a huge
amount of new data will be available, in particular for the
three-dimensional observations (via redshift) of galaxy distribution
and CMBR anisotropies.  Cosmology therefore is based more on
observational and testable grounds than ever in the past. On the other
hand, modern statistical physics can be able to provide a new general
framework for the understanding of cosmic structures, both from the
phenomenological and the theoretical points of view. It is important
to notice that this new approach includes, as a particular case, the old
analytical liquid-like approach (homogeneous mass density fields with
small fluctuations), but it is also able to shed light in more complex
cases as matter distribution in the universe seems to be. Here we
briefly illustrate the main points discussing the properties of galaxy
data and standard cosmological models (see \cite{book} for more
details)



We firstly introduce the basic properties of those stochastic systems
(e.g. mass density fields) that can be represented as Stationary
Stochastic Process (SSP), where {\em stationary} refers to invariance
under translation of the spatial statistical properties. The single
realization of such a SSP can be thought to be a particular density
field. In the context of cosmology the requirement of statistical
stationarity and isotropy is justified by the fact that in the study
of the mass distribution in the universe the cosmological principle is
assumed: there are no preferential points or directions in the
universe.  Clearly this principle must be intended in the statistical
sense. This implies that the statistical properties of the mass
distribution inside a sample volume should not depend on the location
of the sample in the universe and on its spatial orientation: such a
condition can be satisfied in both distributions with positive or zero
ensemble average density (i.e. homogeneous or fractal). 

Let us consider the stochastic mass distribution represented by the
microscopic mass density function $\rho(\vec{r})$ and focus on the
case in which it is a discrete particle distribution: The integral of
$\rho(\vec{r})$ over an arbitrary volume gives the number of
particles in such a volume. Considering particles of identical
unitary mass this is also the mass contained in the volume. We may
then write $\rho(\vec{r})=\sum_{i} \delta(\vec{r} -\vec{r}_i)\,,$
where $\vec{r}_i$ is the position vector of the particle $i$ of the
distribution and $\delta(\vec{x})$ is the Dirac delta function.  In
this context the meaning of {\em homogeneity} (or spatial uniformity)
in terms of the spatial average in a single realization of a
stochastic mass distribution can be expressed as follows.  Due to the
usual assumption of ergodicity of the stochastic mass field, for a
single realization of the mass distribution (i.e. a strictly
non-negative field) the existence of a well-defined average positive
density implies that \cite{book} in $d$ dimensions
\begin{equation}      
 \label{shi1}     
 \lim_{R\rightarrow\infty}      
 \frac{1}{\|S(R;\vec{x}_0)\|}\int_{S(R,\vec{x}_0)}\rho(\vec{r})d^dr    
=\rho_0>0     
\,\,\, \forall \vec{x}_0\,. \,   
\end{equation}  
where $\|S(R,\vec{x}_0)\| \sim R^d$ is the volume of the 
sphere $S(R,\vec{x}_0)$ of radius $R$, centered on an arbitrary point 
$\vec{x}_0$. 

A large scale homogeneous stochastic mass distribution is in general
 characterized by the presence of structured (i.e. correlated)
 fluctuations around the average density, which can be of different
 nature depending on the ensemble statistical properties of the field.
 In order to characterize spatially these structures it is convenient
 to use the two-point correlation function. For example one may
 consider the {\em reduced} two-point correlation function defined as 
$C_2(r_{12})=\left<\left({\rho}(\vec{r}_1)-\rho_0\right)
 \left({\rho}(\vec{r}_2)-\rho_0\right)\right> $ (where
 $\langle.\rangle$ is the ensemble average symbol and 
$r_{12}=|\vec{r}_1-\vec{r}_2|$), which is the main
 function used to study and characterize spatial correlations between
 fluctuations around the positive average value.  From the above
 definitions we see that $C_2(r_{12})$ measures the spatial {\em
 memory} of mass density fluctuations on the scale $r_{12}$.  In order
 to characterize through a single number the persistence of
 correlations in the fluctuation field, the concept of {\em
 correlation length} $r_c$ has been introduced, for example by:
$r_c^2=\frac{\int d^d r\,r^2 C_2(r)}  
{\int d^d r\,C_2(r)}\,.  $
The concept of correlation length is useful to distinguish two
important cases of large $r$ behavior of $C_2(r)$ (see
Fig.\ref{subpoiss}): (i) $C_2(r)\sim r^{-\gamma}$ with $0<\gamma <d$,
and (ii) $C_2(r)\sim \exp (-r/r_c)$ ($r_c$ is a measure of the
correlation length) at large enough $r$. The case (i) is typical of a
thermodynamic system (e.g. the mass density of a fluid at the
liquid-gas critical point) at the point of a second-order phase
transition and the main feature is the presence of fluctuation
structures of all sizes (i.e. the fluctuation field around the
positive average has fractal features), while (ii) is the ordinary
behavior of it far from criticality : for $r\gg r_c$ we have
uncorrelated Poisson-like fluctuations (and for this reason the 
system is called {\em substantially Poisson}).
\bef \epsfxsize 5cm \epsfysize 5cm \centerline{
\epsfbox{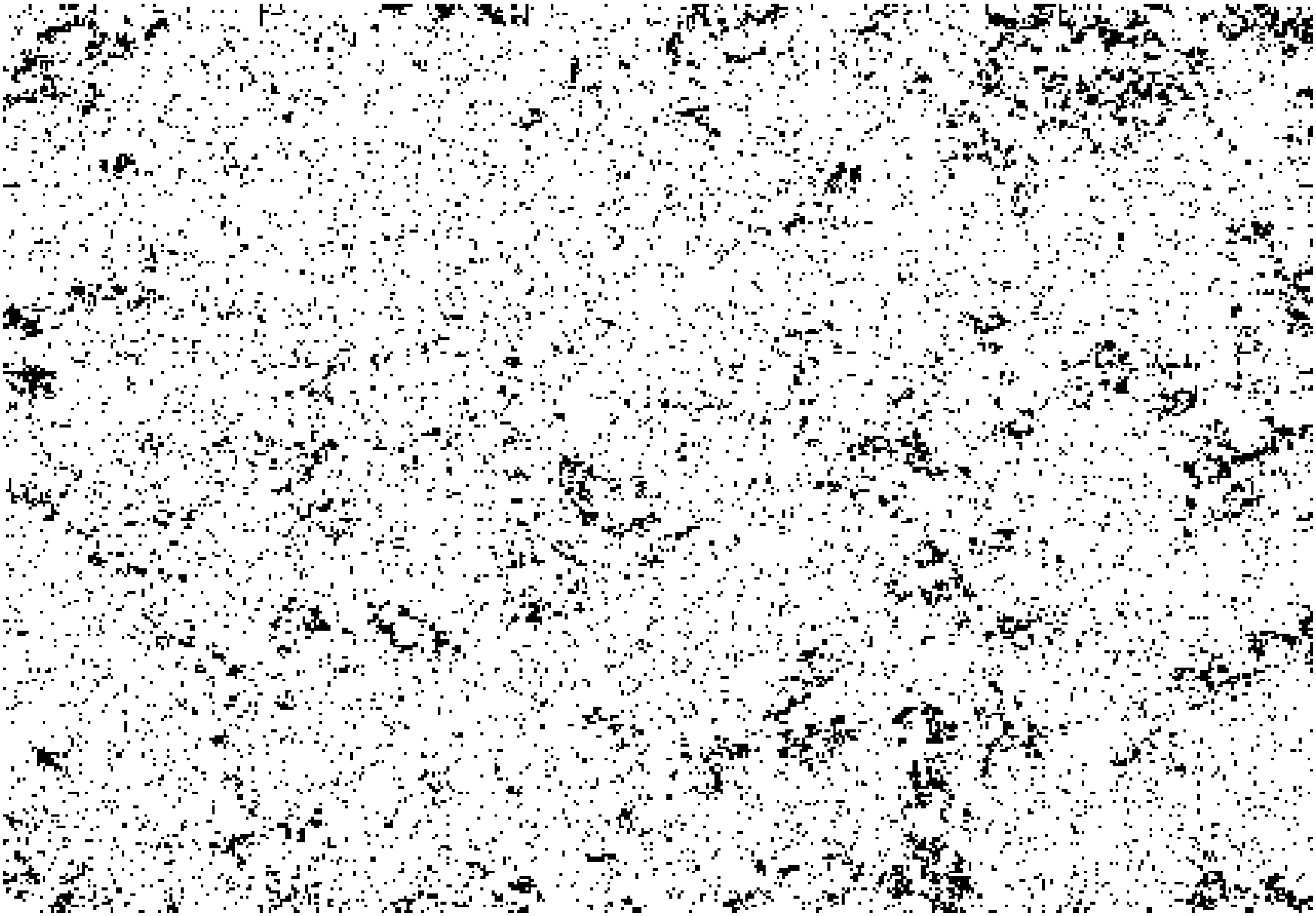} \epsfxsize 5cm \epsfysize 5cm
\epsfbox{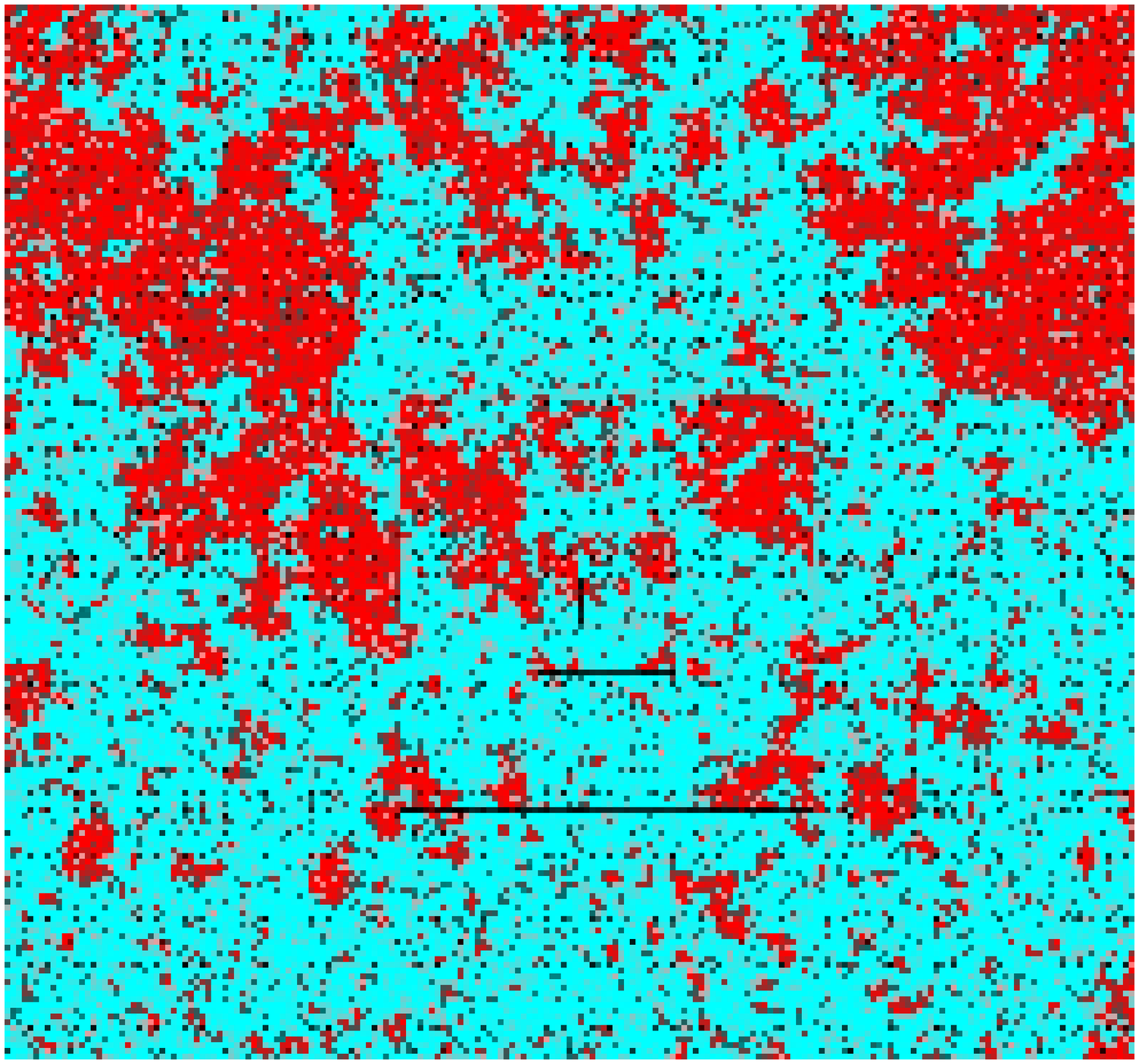}}
\caption{{\it Left panel} Substantially Poisson system: 
Statistically stationary and isotropic particle distribution with positive small-scale 
correlations.  
{\it Right panel:} Homogeneous critical system: 
the system is characterized by a positive background density with large scale and power law
correlations of density fluctuations (in the picture positive fluctuations are dark and 
negative ones are clear)
There are clusters of fluctuations of a fixed sign of     
all sizes and the correlation length diverges $r_c\rightarrow \infty$.  
The absence of an intrinsic characteristic scale 
is shown by the fact that at different ``zooms'' the system looks  
the same, i.e. the fluctuations field (and not the whole filed) 
is self-similar and has {\em fractal} 
features.} 
\label{subpoiss} 
\eef

While the correlation length is the standard tool used to classify long
or short range correlations in homogeneous stochastic density fields, it is 
not suitable to detect the presence of a sort of long-range order.
To this aim one has to focus  on the large
scale behavior of integrated mass fluctuations.  Let
$M(R)=\int_{S(R)}{\rho}(\vec{r}) d^dr$ be the stochastic mass included 
in the sphere $S(R)$ of radius $R$. 
Fluctuations of this quantity are measured by the its variance $\Sigma^2(R) =
\langle M(R)^2 \rangle - \langle M(R) \rangle^2$.    
One can show that for a homogeneous mass density field
with a well defined positive average 
$\rho_0$, the statistical counterpart of the homogeneity condition 
(Eq.\ref{shi1}) is 
$\lim_{R \rightarrow \infty}   
\Sigma^2(R)/\langle M(R) \rangle^2 = 0$.
\bef 
\epsfxsize 6cm
\epsfysize 5cm
\centerline{\epsfbox{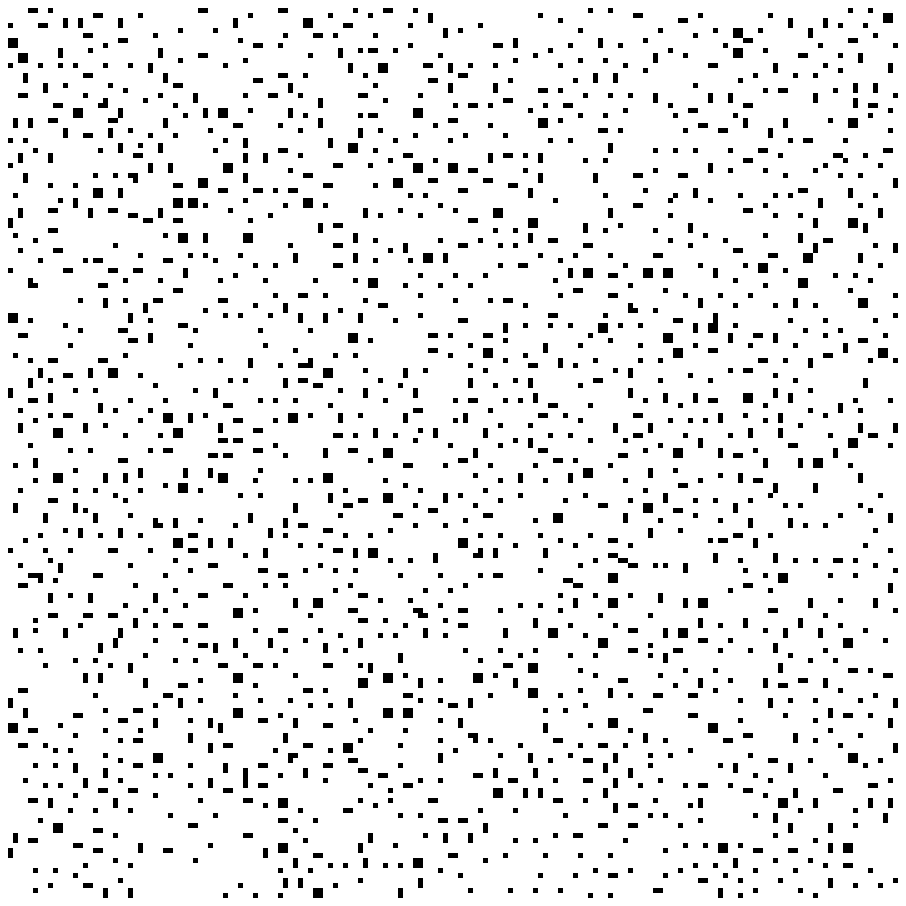}
\epsfxsize 5cm
\epsfysize 5cm
\epsfbox{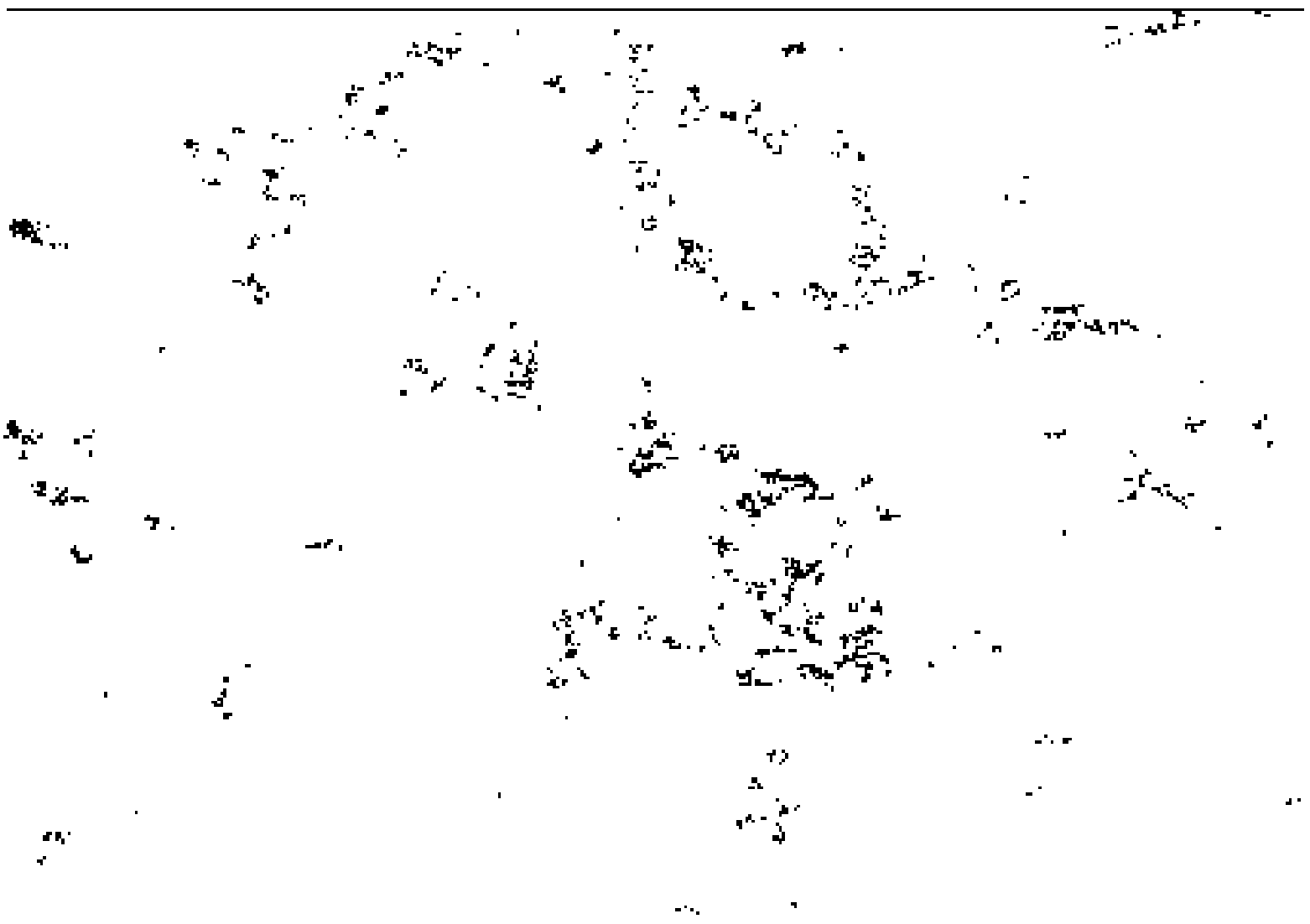}} 
\caption{ {\it Left panel:} Super-homogeneous configuration:  This
is a realization of the One-Component Plasma.
This is a projection of thin slices of
three dimensional distributions. (From \cite{lebo}). 
{\it Right Panel:} 
Fractal distribution with $D=1.47$ in the two dimensional 
 Euclidean space. In this case there are structures and voids of all sizes} 
\label{glass} 
\eef 
Distributions (i) with infinite correlation length (critical systems)
have $\Sigma^2(R) \sim R^{d+\gamma}$ (with $0<\gamma <d$).
Systems (ii) with finite correlation length (i.e. substantially Poisson) 
present $\Sigma^2(R) \sim R^d$.
Finally there is a special case
of distribution characterized (iii) by a sort of long-range order
(super-homogeneous) in the density fluctuations for which
$\Sigma^2(R)$ increases even more slowly than case (ii), i.e.
$\Sigma^2(R) \sim R^{d-\alpha}$ with $0<\alpha\le 1$. The case
$\alpha=1$ is the limiting behavior for any stochastic mass
distribution and it is due to a precise balance between positive and
negative spatial correlations someway similar to what happens in an
ordered lattice particle distribution.

Fractal geometry has allowed us to classify and study a large variety
of structures in nature which are intrinsically irregular and
self-similar \cite{man82}.  Let us limit the discussion to
statistically stationary and isotropic mass fields. The so-called {\it
fractal} dimension $D$ is the most important concept introduced to
describe these intrinsically irregular systems.  Basically it measures
the large scale average logarithmic rate of increase of the ``mass''
$\left<M(R)\right>_p$ around an arbitrary point of the system (for
this reason it is said {\em conditional} mass) with the size $R$ of
the volume in which it is measured: i.e.  $\left<M(R)\right>_p\sim
R^D$.  In any case $0<D\le d$. $D=d$ is found in all previous
homogeneous systems where $\rho_0>0$ is well defined.  For $D<d$
the conditional mass density seen in average by a point
of the system goes to zero in a slow power law way, i.e. the system is
{\em asymptotically empty}.  Note that, however, $D<d$ together with
the spatial statistical stationarity of fluctuations imply 
that the mass field is wildly fluctuating at all scales.  
The concept of fractal dimension is important even for large scale homogeneous
systems (e.g. possibly the galaxy distribution) to characterize {\em locally}
the small scales region of strong clustering where the average density
fluctuations are much larger than $\rho_0>0$. The scale beyond which 
fluctuations are smaller than $\rho_0$ is called {\em homogeneity scale}
and marks the cross-over between the {\em locally} fractal and the homogeneous
behaviors.

Summarizing, we can frame stationary stochastic mass fields in two
classes: (H) homogeneous (i.e. $\rho_0>0$) or (F) fractal 
(asymptotically empty).  
In the (H) class, the less asymptotically fluctuating systems 
are the super-homogeneous ones which are
characterized by a sort of long-range order \cite{lebo}; then 
larger fluctuations are found in the {\em substantially Poisson}
systems with only small scale mainly positive correlations. 
Still larger fluctuations are found in homogeneous 
critical systems characterized
by slow power-law correlations and a positive average density. 
Finally, we have the class (F) of self-similar mass
distributions which are intrinsically inhomogeneous at all scales 
(the conditional average density decreasing
as a slow power-law with the scale).

 
In order to connect this analysis to cosmology we have to consider that 
from one side there are the striking observations
of the three-dimensional galaxy distribution which
have shown a network of large scale irregular 
clusters, filaments and voids (see Fig.\ref{galaxy}). 
\bef 
\epsfxsize 9cm
\centerline{\epsfbox{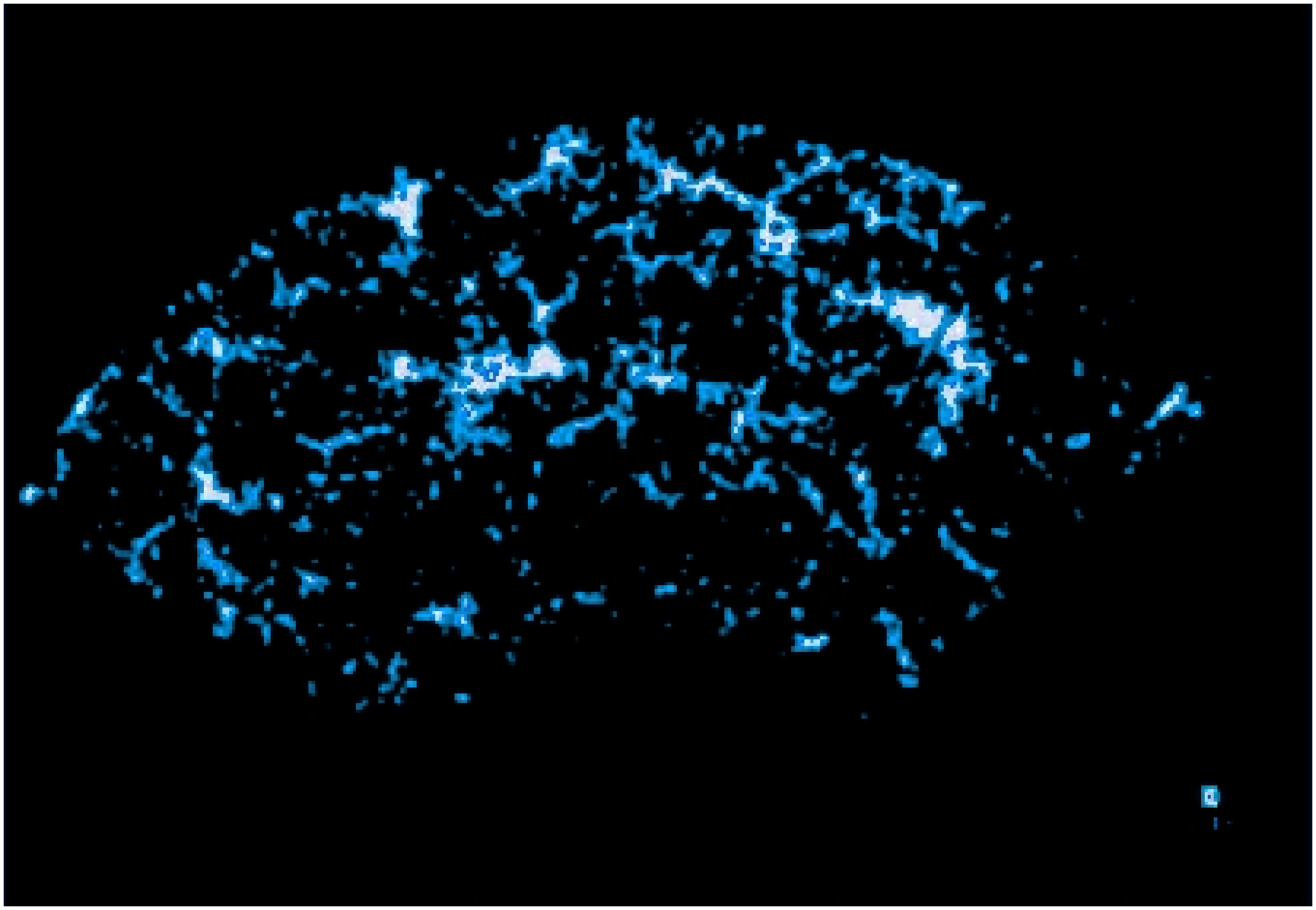}}
\caption{ This is a two dimensional slice of thickness 5 Mpc of a
volume limited sample extracted from the first data release of the
Sloan Digital Sky Survey. The dimension of this sample is
$600 \times 300$ Mpc/h. The color scheme is proportional to the 
local density of galaxies. In this case there
is no bias neither due to luminosity selection effects, nor to
orthogonal projection distortions: Structures of galaxies of hundreds
Mpc are still visible.  The small square at the
bottom right has a side of 5 Mpc, the typical clustering length
according to the standard approach to the description of 
galaxy correlations.
} 
\label{galaxy} 
\eef 
Actually on small but relevant scales (at least up to $20$ Mpc/h)
there is a general agreement that galaxy structures show fractal-like
large fluctuations with power-law correlations \cite{slmp98,rees99}.  
On the other side we find that all standard cosmological models 
compatible with Friedmann metric (e.g. the cold and the 
hot dark matter models) 
predict that the cosmological density field
becomes super-homogeneous \cite{book} (i.e., in view of the previous
discussion, belonging to the class of less fluctuating stochastic mass
distributions) on scales $r\gtapprox 100$ Mpc/h.
Indeed such a feature should be supported by
observations of CMBR anisotropies \cite{wmap}: particularly the large
angular scale anisotropies should trace the super-homogeneous part of
the distribution \cite{book}.  The main theoretical problem is how to
relate the super-homogeneous density field observed in the CMBR to a
fractal-like distribution of galaxies and clusters on smaller
scales. These are two different observations which have to be put
together in a coherent theoretical framework.  In this context the
central question concerns the extension of the fractal behavior: 
how large are the homogeneity scale and 
the largest non-linear (or strongly clustered) structures in the universe?  
This question introduces the next one: how can such large scale
structures of tens or even hundreds of Mpc have formed in the Hubble time
$T_H$ (time scale of the universe)? The typical peculiar velocity of a
single galaxy is about $500$ km/s and in a time $T_H \sim 15$ Giga-year
it has traveled, in average, for few Mpc. Thus we would expect not to
see significant strong clustering on scales larger than 10 Mpc
\cite{saslaw}. The explanation of the formation of these structures
represent the main challenging problem in modern cosmology.

We warmly thank M. Joyce \& L. Pietronero for fruitful
collaborations. FSL acknowledges the support of a Marie Curie Fellowship
HPMF-CT-2001-01443.

\end{document}